# Distribution uniformity of laser-accelerated proton beams*

Jun-Gao Zhu (朱军高), Kun Zhu (朱昆)[1], Li Tao (陶立), Xiao-Han Xu (徐筱菡), Chen Lin (林晨), Wen-Jun Ma (马文君), Hai-Yang Lu (卢海洋), Yan-Ying Zhao (赵研英), Yuan-Rong Lu (陆元荣), Jia-Er Chen (陈佳洱), Xue-Qing Yan (颜学庆)[2]

State Key Laboratory of Nuclear Physics and Technology, Peking University, Beijing 100871, China

Compared with conventional accelerators, laser plasma accelerators can generate high energy ions at a greatly reduced scale, due to their TV/m acceleration gradient. A compact laser plasma accelerator (CLAPA) has been built at the Institute of Heavy Ion Physics at Peking University. It will be used for applied research like biological irradiation, astrophysics simulations, etc. A beamline system with multiple quadrupoles and an analyzing magnet for laser-accelerated ions is proposed here. Since laser-accelerated ion beams have broad energy spectra and large angular divergence, the parameters (beam waist position in the Y direction, beam line layout, drift distance, magnet angles etc.) of the beamline system are carefully designed and optimised to obtain a radially symmetric proton distribution at the irradiation platform. Requirements of energy selection and differences in focusing or defocusing in application systems greatly influence the evolution of proton distributions. With optimal parameters, radially symmetric proton distributions can be achieved and protons with different energy spread within ±5% have similar transverse areas at the experiment target.

**PACS**: 41.75.Jv, 41.85.Lc, 87.56.bd        **DOI**: 10.1088/1674-1137/41/9/097001

*Supported by National Natural Science Foundation of China (Grant No. 11575011，No.61631001) and National Grand Instrument Project (2012YQ030142).

[1] E-mail: zhukun@pku.edu.cn

[2] E-mail: x.yan@pku.edu.cn

## 1. Introduction

Intense laser pulses impinging on solid targets can produce high-energy particle beams [1]. The interaction between an ultra-intense laser and plasma has the potential to reach TV/m accelerating electric field gradient [2] and achieve compact and economical acceleration.

Thousands of patients with tumors are treated by proton radiotherapy each year now. Proton (and heavy ion) cancer therapy equipment based on conventional accelerators, is installed at hospitals, usually has a large scale. In recent years, many theoretical simulation works on laser plasma acceleration have achieved encouraging results [3–7], which has stimulated proposals to use laser acceleration as a new and attractive technology for radiation therapy [8].

In contrast to the ion beams with small energy spread delivered from conventional accelerators, laser-based accelerators inherently yield proton beams with broad energy spectra and large angular divergence [9,10]. These drawbacks make it difficult to obtain uniform proton distribution at the experiment target, which thwarts applications of proton cancer therapy.

A narrow energy spread is necessary in many applications. Many kinds of devices have been tried and tested to cope with wide energy spectra, such as permanent magnet quadrupole lenses [11,12], solenoid magnets [13–15], laser triggered micro-lenses [16], bending magnets [17], a set of dipole magnets [18–22] or a combination of magnets [23–25].

A compact laser plasma accelerator (CLAPA), based on the RPA-PSA mechanism [26,27] or other acceleration mechanisms [28], is under construction at Peking University. A 200 TW, 5 Hz Ti:sapphire laser system based on the double Chirped Pulse Amplification (CPA) technology has been installed. The laser system delivers pulses with 5 J energy, 25 fs duration, a focal spot of about 4μm diameter (full-width at half-maximum (FWHM)) and contrast of $10^{-10}$ at 100 ps. Many investigations will be carried on CLAPA: research on the physical mechanism of laser-plasma acceleration, transmitting of beams with broad energy spectra and large angular divergence, preparation and application of self-supporting ultra-thin targets, ultrathin carbon nanotube foam (CNF) targets [29], metal targets [30], application research in medicine and radiotherapy [9,31], inertial confinement fusion (ICF) [32], astrophysics etc.

To realize these investigations with this laser system, a beam line is designed with magnetic devices to transmit proton beams with energy of 1~44 MeV, energy spread of 0~±5%, and $10^{6-8}$ protons per pulse, to satisfy the requirements of different experiments, especially those involving biomedical irradiation. The transport of proton beams is simulated mainly around the energy of 15 MeV.

## 2. Beam line

The beam line consists of three sections: a Collection System, namely a quadrupole-triplet lens and a quadrupole-doublet lens which help to collect high energy particles, an Energy Selection System including a 45° bending magnet and two slits, and an Application System, namely another quadrupole-triplet lens or quadrupole-doublet lens to focus and deliver particles to the experiment target with an adjustable final beam size (Fig. 1(a)). The specific parameters for this beamline are presented in Fig. 1(c).

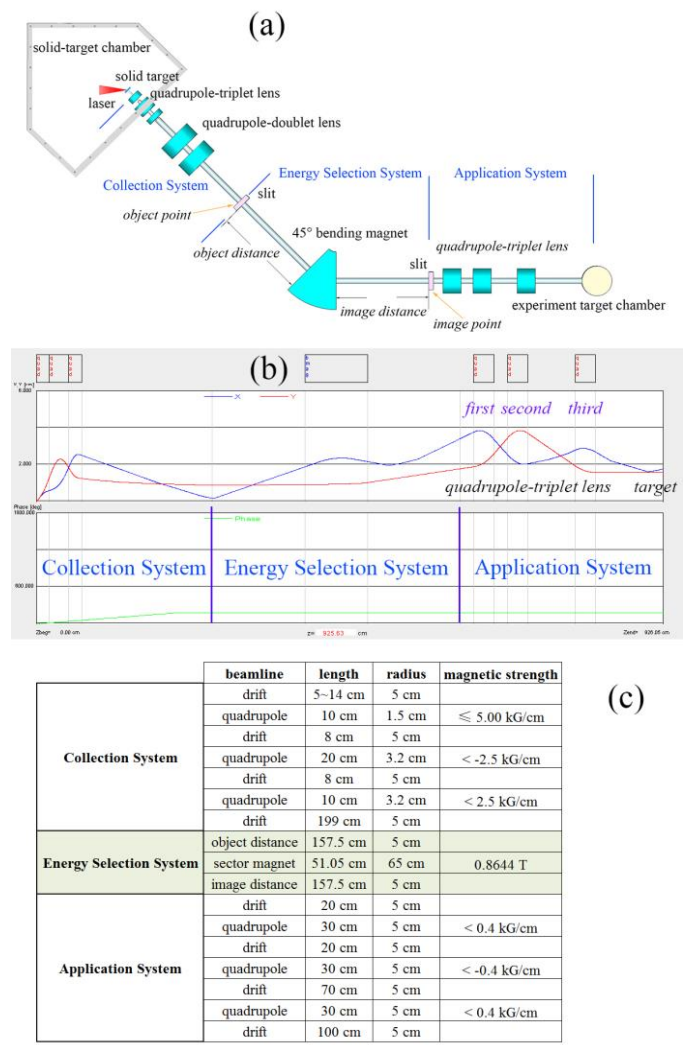

Fig. 1. (a) Schematic of beam line layout. (b) Simulation transport result of 15 MeV proton beam with energy spread of ±5%. (c) Specific parameters for three sections of the beamline: lengths of each part and their apertures, magnetic strengths of the quadrupoles and sector magnets.

### 2.1 Collection system

An aperture is used to screen particles with big divergence angles after the proton beam is generated by laser acceleration. The distance between the laser target and aperture is selectable as needed, ranging from 5 cm to 14 cm in our case. In the presence of a proper aperture, the proton beams drift into the quadrupole-triplet lens with a divergence angle of ±50 mrad, transverse emittance of 0.25 πmm.mrad and current of $1 \times 10^{6\sim8}$ proton/ pulse.

The triplet lens is inserted into a chamber to close to the target, collecting protons with energy up to 19 MeV. All magnetic strengths at the pole face are not more than 0.75 T and magnetic field gradients can be adjusted to move beam waists in the X and Y directions. For instance, for 15 MeV protons, the magnet field gradients may be respectively 5.00, -2.00 and 2.18 kG/cm when the distance between laser target and the lens is 5 cm, and the waist in the X direction is 199 cm from the triplet lens. A doublet lens outside the target chamber will work to assist focusing when the energy is higher.

### 2.2 Energy selection system

Many applications require a high energy proton beam with a small energy spread. The proton beam is focused by the collection system to form a beam waist in the X direction at the object point of the bending magnet in the energy selection system, where a slit is placed to roughly remove particles with a large energy spread. Then the beam is analyzed by a 45° bending magnet without edge angles and converged to form a beam waist at the image point in the X direction. At this point, protons with different energies have been separated in the X direction and a slit is set to keep protons within a ±5% energy spread (Fig. 1(a)).

The radius of the bending magnet is designed as 65 cm and the magnetic field is about 0.8644 T. The airgap width is 7 cm.

### 2.3 Application system

A triplet lens or doublet lens with 5 cm radius is put 20 cm away from the image point and can be adjusted to move the beam waist in different experiments. For example, the magnet field gradients are respectively about 0.32, -0.33 and 0.27 kG/cm in Fig. 1(b). The length of each quadrupole is 30 cm. The distance between the first and second quadrupoles is 20 cm. The third quadrupole is set 70 cm away from the second one. The drift space between the third quadrupole and final experiment target is 100 cm.

Referring to theoretical simulation and studies in the literature [3,33], the initial transverse radius of the beam spot is set as 0.005 mm, which is almost identical to that of the laser spot. The initial pulse duration is set as 20 ps. Correspondingly, the initial longitudinal length is about 1 mm.

The precise energy spectrum of the beam may vary under different acceleration conditions, while it shows a Gaussian-like spectral shape in some energy ranges in experiments [34]. Hence the initial beam distribution is set as a 6D waterbag. Program Track is used to simulate beam transmission with 10001 macro particles [35].

The transport of 15 MeV proton beams with ±5% energy spread, divergence angle of ±50 mrad and transverse emittance of 0.25 πmm.mrad is simulated in this paper. The evolution of the beam envelope is shown in Fig. 1(b).

### 3. Beam waist in the Y direction and uniformity of particle density distribution

Uniformity of particle density distribution is very important in many applications, such as proton cancer therapy requiring specific dose deposition in organic tissue. Nonuniform proton distributions are major drawbacks in plenty of experiments and simulations [18]. Up to now, the study of distribution uniformity of protons delivered in a beam line designed for laser acceleration is scarce.

After protons come out of the collecting lens, they drift a long distance in the Y direction before entering the application system. It is better to converge to a beam waist in the Y direction to control the envelope during this drift. After study, it is found that the location of the beam waist in the Y direction has a crucial influence on the proton distribution.

If the first quadrupole in the application system focuses in the X direction, a quadrupole-triplet lens is needed to deliver all protons to the final experiment target.

#### 3.1 Beam waist in the Y direction inside the bending magnet

When the location of the beam waist in the Y direction is in the middle of the bending magnet, the envelope in the Y direction inside the bending magnet will be at a minimum, namely about a radius of 1 cm for ±5% energy spread. With proper magnetic strengths in the application system, the envelope sizes in the X and Y directions can be close at the final experiment target. At this point, the transport efficiency of protons with ±5% energy spread can be 100%, while the density distributions of the protons are extremely radially asymmetric (Fig. 2). Moreover, delivering with the same beam line, beam spot sizes of ±1% and ±5% energy spread have remarkable distinctions. The process of these results is shown in Fig. 3.

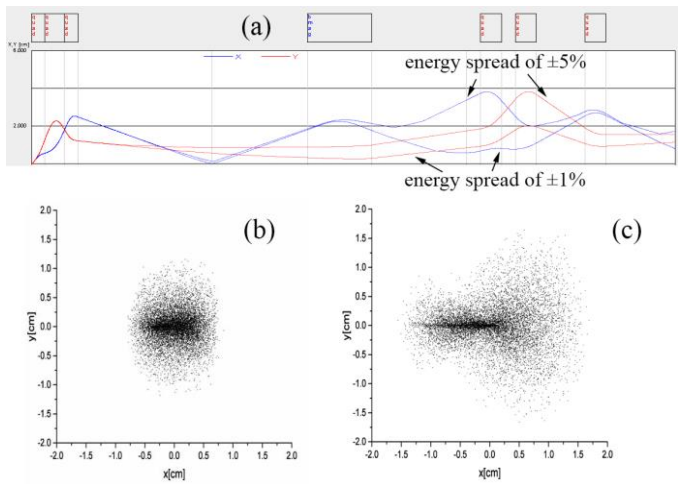

Fig. 2. (a) Envelopes of proton beams contracting ±5% energy spread with ±1% energy spread delivered with the same beam line when the location of the beam waist in the Y direction is inside the bending magnet. (b) Distribution of protons with ±1% energy spread at the experiment target. (c) Distribution of protons with ±5% energy spread at the experiment target.

After analysis by the bending magnet, protons with different energies have been separated at the image point in the X direction. At this point, the distribution of protons with ±5% energy spread is as shown in Fig. 3(a). The vertical axis indicates the energy difference from 15 MeV.

The protons then enter the application system and are focused by a quadrupole-triplet lens. Figure 3(b) shows the proton transverse distribution 20 cm away from the first quadrupole of the lens. After focusing in the X direction, defocusing in the Y direction in the first quadrupole and 20 cm drift, the total envelopes shrink in the X direction and enlarge in the Y direction.

Figure 3(c), (d), (e) and (f) shows the proton transverse distributions at the exit of the second quadrupole, 20, 40 and 70 cm away respectively. It shows the envelopes shrink in the Y direction and enlarge a little in the X direction. But most protons with large energy spread still shrink in the X direction.

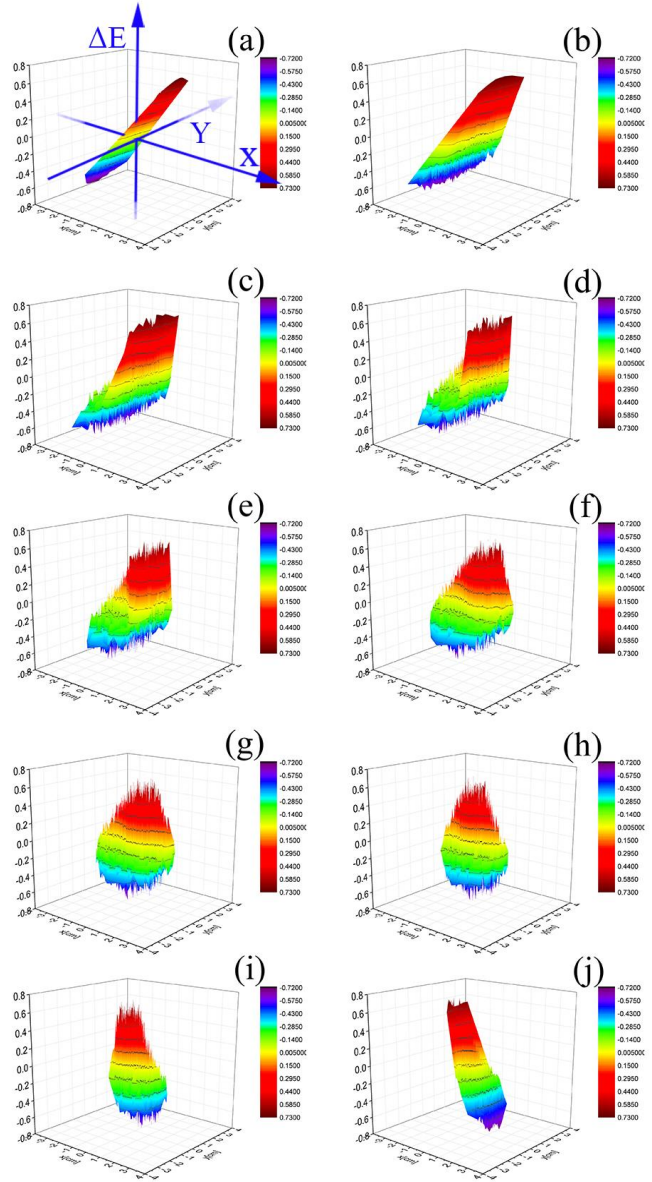

Fig. 3. Evolutions of proton transverse distributions in the application system when the location of the beam waist in the Y direction is inside the bending magnet. All figures have the same coordinate system. The horizontal plane is the XY plane. The vertical axis indicates the energy difference from 15 MeV. (a) Protons at the image point in the X direction. (b) Protons 20 cm away from the first quadrupole. (c) Protons at the exit of the second quadrupole. (d) Protons 20 cm away from the second quadrupole. (e) Protons 40 cm away from the second quadrupole. (f) Protons 70 cm away from the second quadrupole. (g) Protons at the exit of the third quadrupole. (h) Protons 30 cm away from the third quadrupole. (i) Protons 60 cm away from the third quadrupole. (j) Protons at final experiment target, namely 100 cm away from the third quadrupole.

The magnet field gradients in the first and second quadrupoles are respectively 0.317 and -0.332 kG/cm. Figure 2 shows that the envelope of protons with ±5% energy spread in the X direction in the second quadrupole is only about half of that in the first quadrupole, hence protons with large energy spread get more focused in the first quadrupole and less defocused in the second quadrupole in the X direction, as the magnet field gradients are close. For the same reason, protons with energy spread less than ±1% have a larger envelope in the second quadrupole and get more defocused. Envelopes and focusing distinctions in the first and second quadrupoles lead to different evolutions of distributions of protons with different energy spreads.

Figure 3(g), (h), (i) and (j) shows the proton transverse distributions at the exit of the third quadrupole, 30, 60 and 100 cm away respectively. The total envelopes change little in the Y direction and shrink in the X direction. Protons with large energy spread in Fig. 3(f) are still focusing and are more focused in the X direction in the third quadrupole. These protons then pass the beam waist and defocus in the X direction. Finally, protons with different energies have different beam spot areas at the experiment target and proton distributions are radially asymmetric.

### 3.2 Beam waist in the Y direction in front of the bending magnet

If the beam waist in the Y direction is positioned before the object point of bending magnet, it will increase the diverging angle in the Y direction and lead to a bigger envelope (Fig. 4(a)). Proton beams with different energy spread within ±5% delivered with the same beam line have close distribution areas, and the proton distributions at experiment target become more uniform (Figs. 4(b), (c), (d) and (e)). The transport efficiency of protons with ±5% energy spread is about 95%. If the beam waist in the Y direction is posed a little behind the object point, the losses of protons will be less, but the distribution uniformity gets worse (Fig. 8(a)).

### 3.3 Beam waist in the Y direction behind the bending magnet

If the beam waist in the Y direction is located behind the bending magnet, the uniformity of the proton distribution gets worse (Fig. 5), as a smaller diverging angle in the Y direction has a greater impact and it is not easy to disperse these protons.

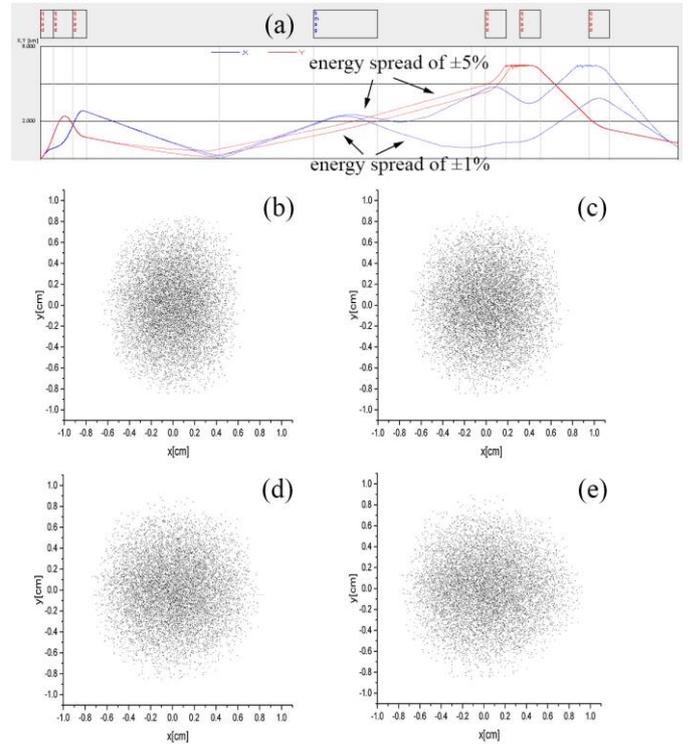

Fig. 4.   (a) Envelopes of proton beams contracting ±5% energy spread with ±1% energy spread delivered with the same beam line when the beam waist in the Y direction is positioned before the object point of the bending magnet. (b) Final distribution of protons with ±1% energy spread. (c) Final distribution of protons with ±3% energy spread. (d) Final distribution of protons with ±4% energy spread. (e) Final distribution of protons with ±5% energy spread.

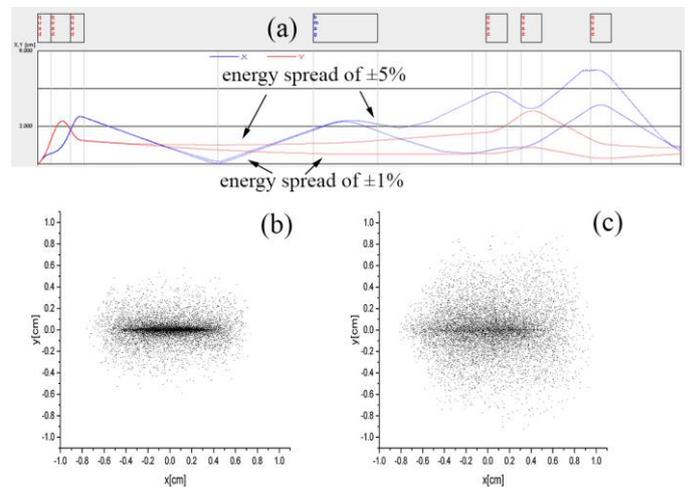

Fig. 5. (a) Envelopes of proton beams contracting ±5% energy spread with ±1% energy spread delivered with the same beam line when the beam waist in the Y direction is positioned behind the bending magnet. (b) Final distribution of protons with ±1% energy spread. (c) Final distribution of protons with ±5% energy spread.

### 3.4 No beam waist in the Y direction before entering the application system

If the proton beam diverges in the Y direction before entering the quadrupoles in the application system, the final proton distributions are not radially symmetric (Fig. 6).

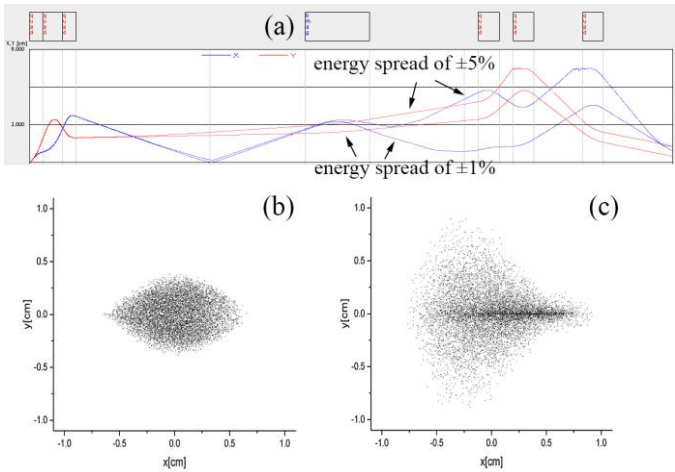

Fig. 6. (a) Envelopes of proton beams contracting ±5% energy spread with ±1% energy spread delivered with the same beam line when the proton beam diverges in the Y direction before entering the application system. (b) Final distribution of protons with ±1% energy spread. (c) Final distribution of protons with ±5% energy spread.

### 4. Envelope control in application system

#### 4.1 Length of drift before arriving at experiment target

The lengths of drift between the third quadrupole and the final experiment target in the above simulations are all 100 cm. If the drift space is shortened, e.g. to 34 cm, the radius of the final beam spot size can be reduced, from 0.92 cm to 0.37 cm (Fig. 7).

In order to improve distribution uniformity, in this situation the best beam waist position in the Y direction is a little before the bending magnet (Fig. 7(b)). Hence, different drift spaces have different optimal positions of the beam waist in the Y direction.

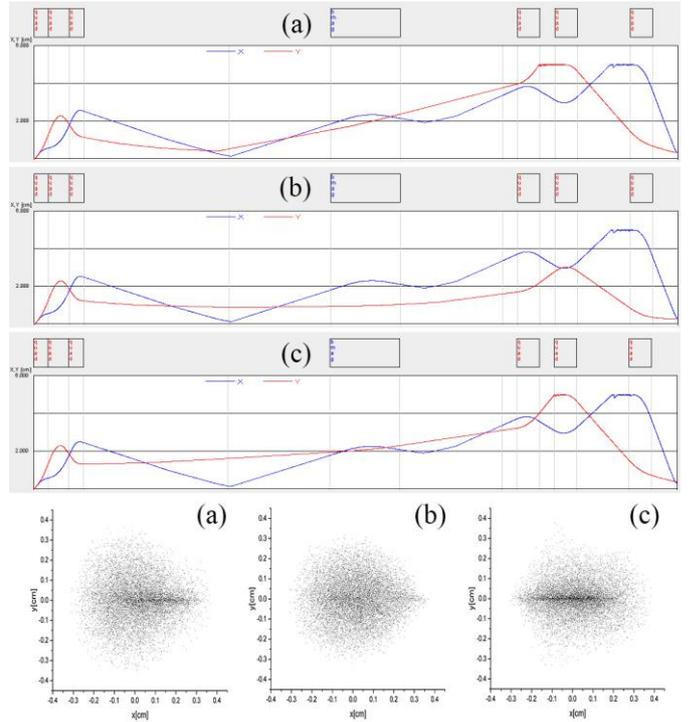

Fig. 7. Envelopes of proton beams with ±5% energy spread and final proton distributions when the length of drift before the experiment target is 34 cm. (a) The beam waist in the Y direction is positioned before the object point of the bending magnet. (b) The beam waist position in the Y direction is a little before the bending magnet. (c) The proton beam diverges in the Y direction before entering the quadrupoles in the application system.

#### 4.2 Envelope control in the X direction

In the application system, the first and third quadrupoles of the lens focus in the X direction. To control the envelope in the X direction, we may increase the magnetic field strength in the first or the third quadrupole. The final proton distribution areas are both round in Figs. 8(a) and (b), but the magnetic field strength in the first quadrupole is smaller and the envelope in the X direction in the third quadrupole is bigger in Fig. 8(a). The final beam spot is smaller and the proton distribution uniformity is better in Fig. 8(a).

In Fig. 8(b), magnet field gradients in the first and second quadrupoles are respectively 0.317 and -0.336 kG/cm. The envelope of protons with ±5% energy spread in the second quadrupole is only about half of that in the first quadrupole, hence protons with large energy spread are focused more in the first quadrupole and defocused less in the second quadrupole in the X direction. For the same reason, protons with small energy spread have a larger envelope in the second quadrupole and are more defocused. Protons with different energies will then evolve inconsistently, which leads to a bigger beam spot at final target. Hence, a smaller magnetic field strength in the first quadrupole and similar envelope sizes in the first and second quadrupoles are better.

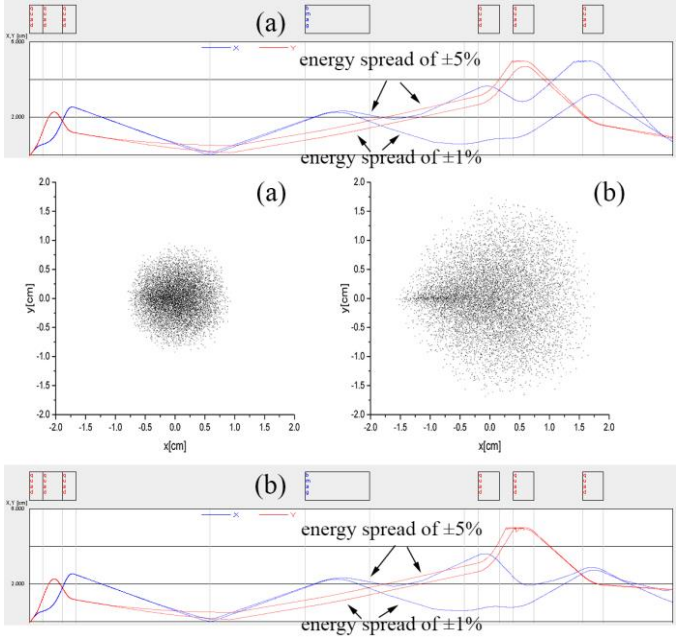

Fig. 8. Envelopes of proton beams contracting ±5% energy spread with ±1% energy spread delivered with the same beam line and circular proton distribution areas at the final target. (a) Magnetic field strength in the first quadrupole in the application system is smaller and the envelope in the X direction in the third quadrupole is bigger. (b). Magnetic field strength in the first quadrupole in the application system is bigger and the envelope in the X direction in the third quadrupole is smaller.

## 5. Changing beam line layout in application system

In the application system, compared with a quadrupole-doublet lens, a triplet lens has stronger focusing ability and reduces particle losses. However, distributions of protons at the experiment target are not radially symmetric when the location of the beam waist in the Y direction is inside the bending magnet (Fig. 2). In the doublet condition, if the first quadrupole in the application system defocuses in the X direction, the final proton distribution can easily be radially symmetric. At this point, a doublet lens is needed, at the cost of losing about 10% of the protons. The reason can be demonstrated below.

Figure 9 depicts the distributions of protons with negligible energy spread at different locations along the beam line without a focusing lens in the application system when the location of the beam waist in the Y direction is inside the bending magnet. It is evident that the positions of most protons in the Y direction change little even after long drift. The reason is that the diverging angle in the Y direction is relatively small and most protons are almost collimated in the Y direction.

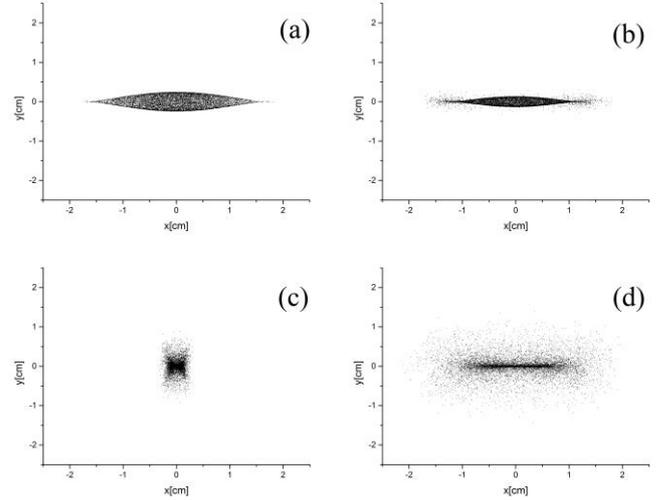

Fig. 9. Proton distributions with energy spread of ±0.0000000001%: (a) before entering the bending magnet; (b) after coming out of the bending magnet; (c) 20 cm away from the image point in the X direction; and (d) 150 cm away from the image point in the X direction without a focusing lens in the application system.

If the first quadrupole defocuses in the X direction, these protons can be easily scattered and the distribution becomes radially symmetric at the final target. A proton beam with ±1% energy spread has uniform distribution when the location of the beam waist in the Y direction is inside the bending magnet (Fig. 10). This can be explained by the proton beam being defocused and dispersed in the first quadrupole in the X direction and focused in the second, then protons being re-distributed and decentralized in the Y direction in the drift space behind the doublet lens. If the first quadrupole focuses in the X direction, the protons cannot be easily scattered (Fig. 2).

As most protons are closer to the X axis than to the Y axis, focusing or defocusing in the Y direction has much less influence on distribution uniformity.

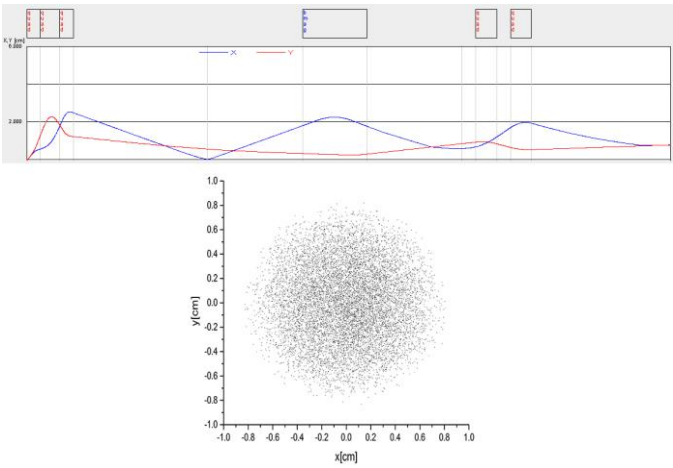

Fig. 10. Envelopes of proton beam with ±1% energy spread and final proton distribution when the first quadrupole in the application system is defocusing in the X direction. The location of the beam waist in the Y direction is inside the bending magnet.

In Fig. 11, the first quadrupole is defocusing in the X direction and the length of drift between the doublet lens and experiment target is 200 cm. As there are losses of protons with large energy spread, the energy spread at the final target is smaller than the initial energy spread. The transport efficiency of protons with ±5% energy spread is about 90%. This efficiency can be 100% if the radius of the doublet lens is 80 mm.

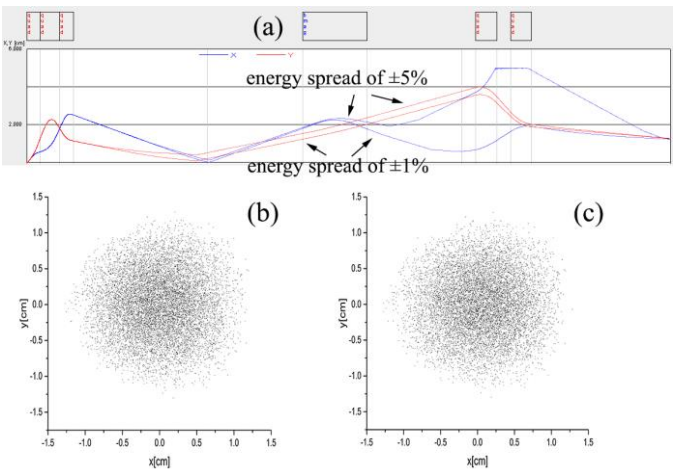

Fig. 11. (a) Envelopes of proton beams contracting ±5% energy spread with ±1% energy spread delivered with the same beam line when there is a quadrupole-doublet lens in the application system and the drift space before the experiment target is 200 cm. (b) Final distribution of protons with ±1% energy spread. (c) Final distribution of protons with ±4% energy spread.

The proton distributions at the final target are relatively uniform and protons with ±1%, ±4% energy spread delivered with the same beam line have close transverse areas (Figs. 11(b) and (c)).

If the length of drift is shortened to 50 cm, the radius of the final beam spot size can be shrunk from 1 cm to 0.4 cm.

## 6. Bending angle of sector magnet

A sector magnet is a conventional component to separate particles with different energies. In practical applications of proton cancer therapy, sector magnets with different bending angles may be needed. The replacements by 60° and 90° sector magnets in this beam line are simulated and the results are given in Fig. 12 and Fig. 13.

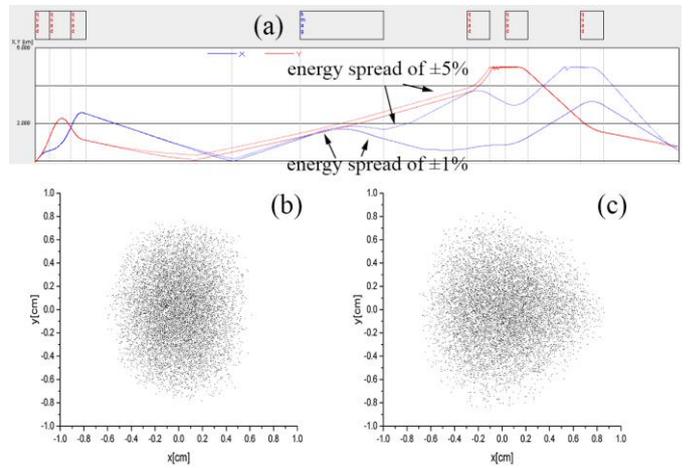

Fig. 12. (a) Envelopes of proton beams contracting ±5% energy spread with ±1% energy spread delivered with the same beam line when the bending angle is 60°. (b) Final distribution of protons with ±1% energy spread. (c) Final distribution of protons with ±5% energy spread.

The bigger the bending angle is, the shorter the total beam line will be. For instance, the total length of beam line with a 90° sector magnet (Fig. 13) is 187 cm shorter than that with a 45° sector magnet with 65 cm bending radius.

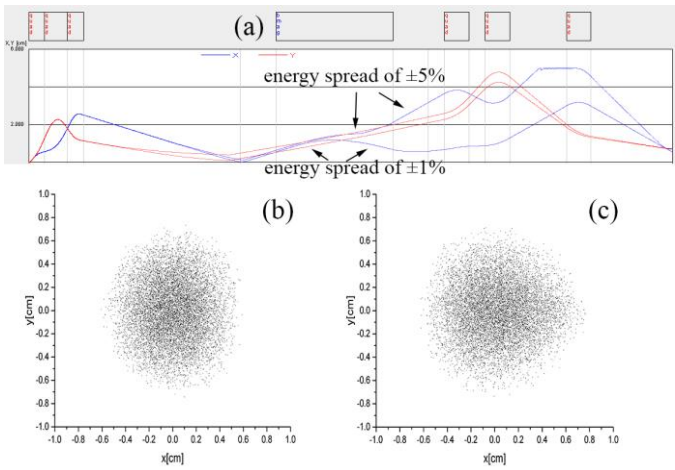

Fig. 13. (a) Envelopes of proton beams contracting ±5% energy spread with ±1% energy spread delivered with the same beam line when the bending angle is 90°. (b) Final distribution of protons with ±1% energy spread. (c) Final distribution of protons with ±5% energy spread.

## 7. Summary

In summary, the beam line of CLAPA is designed with traditional transport elements and able to deliver most protons with energy ranging from 1 to 44 MeV and energy spread within ±5%. The proton distribution at the experiment target can be regulated by many parameters.

It is found that envelope in the Y direction converging to a beam waist before entering the bending magnet is favorable to obtain distributed protons in the Y direction, when the focusing lens in the application system is quadrupole-triplet and the length of drift is 200 cm before the experiment target. Proton beams with different energy spread within ±5% can have close distribution areas. Transport efficiency of protons with ±5% energy spread is about 95%. Protons with different energies should be separated at the image point in the X direction and they have different envelopes and undergo different focusing or defocusing in the first and second quadrupoles in the application system, which determines the evolution of proton distributions. The final beam spot size can be reduced when the drift distance before the experiment target is shortened.

Bending magnets of 60° and 90° in this beam line are simulated and they can reduce the lengths of beam line markedly.

Envelope controls in the X and Y directions both have a critical impact on proton distribution. The envelope in the Y direction has more room for adjusting to get better uniformity. The best position of the beam waist in the Y direction depends on drift distance, beam line layout in the application system, and the angle of the bending magnet.